\documentclass[twocolumn,aps,prl,superscriptaddress,nofootinbib]{revtex4}
\usepackage{graphicx}
\usepackage{amssymb}
\newcommand{\TT}{T_{-}\bar{T}_{-}}
\newcommand{\met}{\not\!\! E_{T}}

\begin{document}
\title{Impact of Single-Top Measurement to Littlest Higgs Model with T-Parity }
\author{Qing-Hong Cao}
\affiliation{Department of Physics and Astronomy, University of California at
Riverside, Riverside, CA 92521}
\author{Chong Sheng Li}
\affiliation{Department of Physics, Peking University, Beijing 100871, China}
\author{C.-P. Yuan}

\affiliation{Department of Physics and Astronomy, Michigan State University, East
Lansing, MI 48824}

\begin{abstract}
We show that a precise measurement of the single-top production cross
section at the Tevatron and the LHC can strongly constrain the model
parameters of the Littlest Higgs model with T-parity. A reduction
in the single-top production rate from the Standard Model prediction
implies new physics phenomena generated by the heavy T-parity partners
of the top quark. We show that the degree of polarization of the top
quark produced from the decay of its heavy T-odd partner ($T_{-}$)
can be utilized to determine the new physics energy scale, and the
mass of $T_{-}$ can be measured from the missing transverse momentum
distribution in the $T_{-}\bar{T}_{-}$ event. 
\end{abstract}
\maketitle
In spite of the great success of the Standard Model (SM) of particle
physics, there is no good understanding why the mass of the SM Higgs
boson is at the weak scale. One extension of the SM, as a low energy
effective theory below the cutoff scale $\Lambda$, is the class of
Little Higgs (LH) models~\cite{Arkani-Hamed:lh} in which the electroweak
symmetry is collectively broken and a weak scale Higgs boson mass
is radiatively generated. At one loop order, the large $\Lambda^{2}$
correction to the Higgs boson mass term induced by the top quark ($t$)
is cancelled by its fermionic partner, and those induced by the electroweak
gauge bosons are cancelled by their bosonic partners. Constraints
from low energy precision data, especially the $\rho$-parameter measurement,
require the symmetry breaking scale of the LH models has to be so
high that the predicted phenomenology has little relevance to the
current high energy collider physics program~\cite{lhprecison}.
To relax the constraints from low energy data, a discrete symmetry
called T-parity was introduced in the Little Higgs models, to warrant
the $\rho$-parameter to be one at tree-level, known as the Little
Higgs models with T-parity\ \cite{lht,Hubisz:2005tx,pheno,todd,msumodel}.
Consequently, the cutoff scale of the model, $\Lambda=4\pi f$, can
be as low as $10\,{\rm TeV}$ and the masses of new heavy resonances,
at the scale of $f$, can be of sub-TeV~\cite{lht}. This type of
models is particularly interesting because it also provides a dark
matter candidate which is the lightest T-odd particle (LTP) $A_{H}$,
the heavy bosonic T-parity partner of photon~\cite{darkmatter}.
Here, we shall focus on the {}``Littlest'' Higgs model with T-parity
(LHT), which is based on an $SU(5)/SO(5)$ nonlinear sigma model whose
low energy Lagrangian is described in detail in Ref.~\cite{msumodel}.
The new particle mass scale $f$ of the model is bounded from below
by low energy precision data to be about $500\,{\rm GeV}$~\cite{Hubisz:2005tx}.
In this work, we will concentrate on the phenomenology associated
with the T-even top partner ($T_{+}$), T-odd top partner ($T_{-}$)
and T-odd partners of the electroweak gauge bosons~\cite{msumodel}.

After the electroweak symmetry breaking, the masses of the T-parity
partners of the photon $(A_{H})$, $Z$-boson $(Z_{H})$ and $W$-boson
$(W_{H})$ are generated as $M_{A_{H}}=\frac{g'f}{\sqrt{5}}(1-\frac{5v^{2}}{8f^{2}}+\cdots)$
and $M_{Z_{H}}\simeq M_{W_{H}}=gf(1-\frac{v^{2}}{8f^{2}}+\cdots)$.
Here, $v$ characterizes the weak scale ($\simeq246$ GeV), and at
tree level the SM-like $W$ and $Z$ gauge boson masses can be expressed
as $M_{W}=\frac{g}{2}v$ and $M_{Z}=\frac{\sqrt{g^{2}+g^{'2}}}{2}v$,
respectively. Because of the smallness of the $U(1)_{Y}$ gauge coupling
constant $g'$, the T-parity partner of the photon $A_{H}$ tends
to be the lightest T-odd particle in the LHT. We note that the mass
of $W_{H}$ is determined by $f$, for the $SU(2)_{YL}$ gauge coupling
constant $g$ and the vacuum expectation value $v$ have been fixed
by the measured values of $W$ and $Z$ boson masses. Hence, if $M_{W_{H}}$
can be directly measured from collider data, then the cutoff scale
of the LHT can be determined. Another way to determine the scale $f$
is to study the T-parity partners of the top quark which is to be
shown below.

\begin{figure}
\includegraphics[clip,scale=0.25]{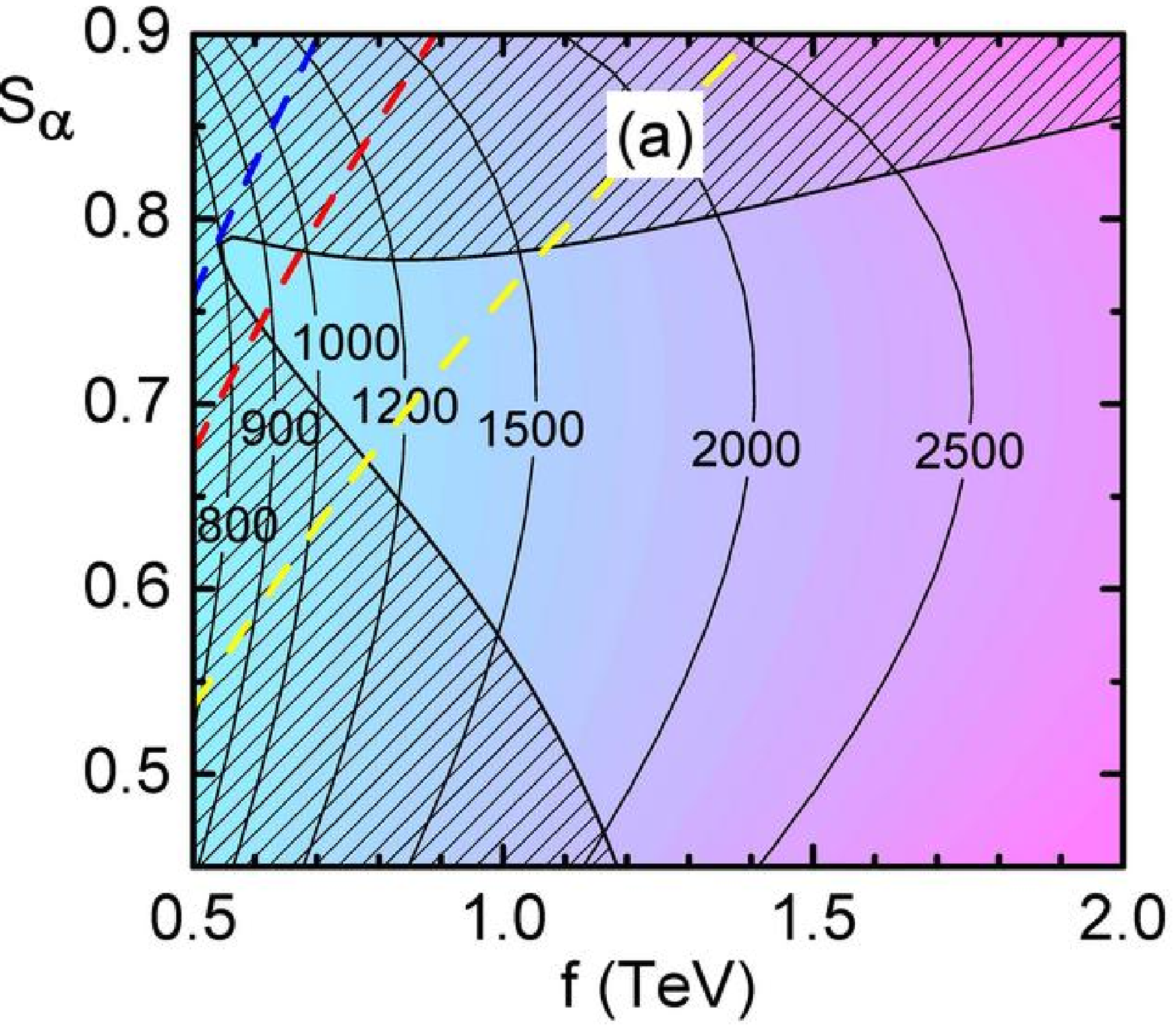}\includegraphics[clip,scale=0.25]{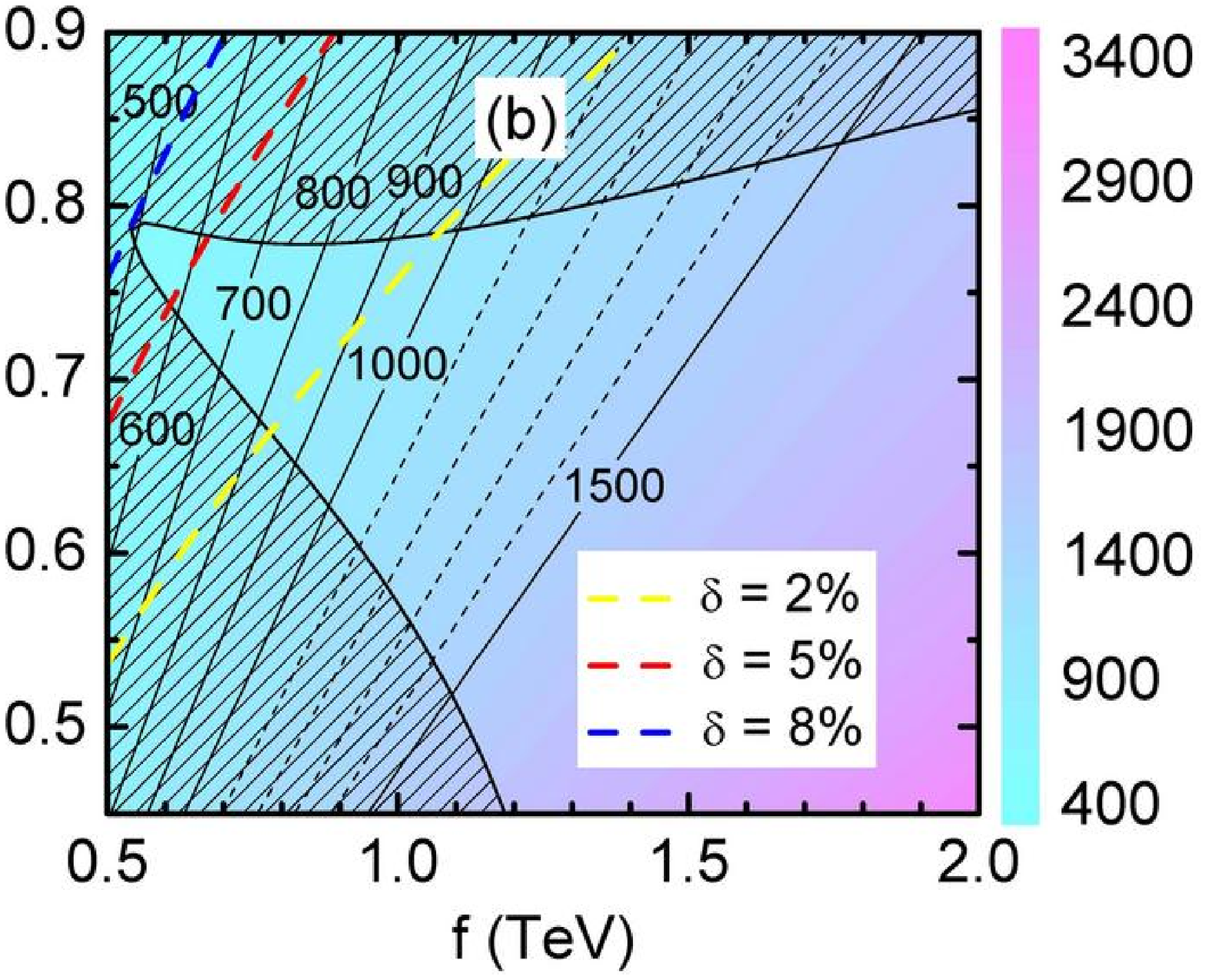}

\caption{Contours of $M_{T_{+}}$ (a) and $M_{T_{-}}$ (b) (in the unit of
GeV) in the plane of $s_{\alpha}$ and $f$. See detailed explanation
in the text.\label{fig:mass}}
\end{figure}

As shown in Ref.~\cite{msumodel}, the mass of top quark is generated
from top-Yukawa interaction Lagrangian which depends on two parameters
of the LHT. In this work they are chosen to be $f$ and the mixing
angle $\alpha$ which describes the amount of mixing among the fermionic
degrees of freedom needed to cancel the quadratic divergence of Higgs
boson mass term at the one loop level. The mass of top quark ($m_{t}$)
has been measured to a good accuracy~\cite{topmass}. Given $m_{t}$
and $v$, we can trade the two parameters $f$ and $\alpha$ by the
masses of the top quark T-parity partners $T_{+}$ and $T_{-}$. Up
to the $O(v^{2}/f^{2})$ corrections, they can be expressed, respectively,
as \begin{eqnarray}
M_{T_{+}} & = & \frac{m_{t}f}{vc_{\alpha}s_{\alpha}},\qquad M_{T_{-}}=\frac{m_{t}f}{v{s_{\alpha}}},\label{eq:mtev-mtod}\end{eqnarray}
 where $s_{\alpha}$ denotes $\sin\alpha$ and is bounded from above
to be less than 0.96 by the unitarity requirement for considering
the $J=1$ partial wave amplitudes in the coupled system of $(t\bar{t},~T_{+}\bar{T}_{+},~b\bar{b},~WW,~Zh)$
states~\cite{msumodel}. Moreover, $\sin\alpha$ cannot be exactly
equal to zero because the {}``collective'' symmetry breaking mechanism
of the LHT only works for a non-zero $s_{\alpha}$, and $M_{T_{+}}$
cannot be larger than the cutoff scale $\Lambda$. If we take the
{}``naturalness'' argument seriously for the Higgs mass corrections,
then $s_{\alpha}$ has to be larger than about 0.1 for $f$ to be
around 1\,TeV~\cite{msumodel}. In Fig.~\ref{fig:mass}, we show
the contours of $M_{T_{+}}$ (left panel) and $M_{-}$ (right panel)
in the plane of $s_{\alpha}$ and $f$.

Due to the mixing between $t$ and $T_{+}$, the couplings of $W_{\mu}^{+}\bar{t}b$
and $W^{+}\bar{T}_{+}b$ are expressed, respectively, as \begin{equation}
V_{tb}\left(i\frac{g}{\sqrt{2}}c_{L}\gamma_{\mu}P_{L}\right)\quad{\rm and}\quad V_{tb}\left(i\frac{g}{\sqrt{2}}s_{L}\gamma_{\mu}P_{L}\right),\label{eq:coupling}\end{equation}
 where $V_{tb}$ is the value of the $(t,b)$ element of the Cabibbo-Kobayashi-Maskawa
(CKM) matrix, $c_{L}=\sqrt{1-s_{L}^{2}}$ with $s_{L}=s_{\alpha}^{2}\frac{v}{f}+\cdots$,
and $P_{L}=\frac{1-\gamma_{5}}{2}$ and $P_{R}=\frac{1+\gamma_{5}}{2}$
are the left-handed and right-handed projection operators, respectively.
In the above expression, the product of $V_{tb}c_{L}$, which is denoted
as $V_{tb}^{eff}$, should be identified with the CKM matrix element
determined from the low energy processes up to the one-loop order.
In the SM, the value of $V_{tb}$ element is constrained by the unitarity
of CKM matrix, which requires its value to be very close to 1 (about
0.999~\cite{pdg}). For simplicity, we will take $V_{tb}=1$ in our
numerical analysis. When the parameter $s_{\alpha}$ varies, the effective
coupling strength of $W^{+}\bar{t}b$, hence the single-top production
rate at the Fermilab Tevatron and the CERN Large Hadron Collider (LHC),
also varies. As $s_{\alpha}\rightarrow0$, it is approaching to the
SM $W^{+}\bar{t}b$ coupling strength. Furthermore, since $|s_{\alpha}|$
is bounded by 1, $c_{\alpha}$ has to be larger than $\sqrt{1-(v/f)^{2}}$
in the LHT; with $f=500$\,GeV, $c_{\alpha}>0.88$. Hence, $V_{tb}^{eff}$
is consistent with the Tevatron measurement on the decay branching
ratio of $t\to bW^{+}$ in the $t{\bar{t}}$ events~\cite{pdg}.
It is also consistent with the most recent measurement of single-top
event rate at the Tevatron: $\left|V_{tb}\right|=1.0_{+0.0}^{-0.12}$~\cite{st-run2}.

Since the strength of the $W_{\mu}^{+}\bar{t}b$ coupling in the LHT
is always smaller than that in the SM, the single-top production rate
at the Tevatron and the LHC will also be smaller than that predicted
by the SM. Hence, the measurement of the single-top production cross
section provides a crucial test to the LHT. The deviations of the
cross sections of the single-top production from the SM predictions
($\delta\equiv\Delta\sigma/\sigma_{{\rm SM}}$) can be expressed in
terms of $s_{\alpha}$ and $f$ as \begin{equation}
\delta=\frac{\sigma_{SM}-\sigma_{LHT}}{\sigma_{SM}}={s_{\alpha}^{4}}\frac{v^{2}}{f^{2}}+\mathcal{O}\left(\left(\frac{v}{f}\right)^{4}\right).\label{eq:wtb-shift}\end{equation}
 For illustration, we show in Fig.~\ref{fig:mass} the constraints
on the parameter $s_{\alpha}$ and $f$ for $\delta=2\%$ (yellow
dashed-line), $5\%$ (red dashed-line) and $8\%$ (blue dashed-line),
respectively. The shadowed region respects the electroweak precision
test (EWPT), with $V_{tb}$ set to be 1, at the $95\%$ confidence
level~\cite{Hubisz:2005tx}. In the same figure we also show the
contours of $M_{T_{+}}$ (left panel) and $M_{T_{-}}$ (right panel).
The pattern of the contour lines can be easily understood from Eq.~(\ref{eq:mtev-mtod}).
We note that the allowed parameter space is strongly constrained when
$\delta$ is small. For example, when $\delta\lesssim2\%$, $f\ge780\,{\rm GeV}$
and $0.67<s_{\alpha}<0.78$; when $\delta\lesssim5\%$, $f\geq600\,{\rm GeV}$
and $0.74<s_{\alpha}<0.85$; and when $\delta\lesssim8\%$, $f\ge550\,{\rm GeV}$.
The above constraints can be translated into non-trivial limits of
$M_{T_{+}}$ and $M_{T_{-}}$, which are summarized in Table~\ref{tab:mass-limie}.

At hadron colliders, single-top events can be produced via three processes:
$s$-channel ($u\bar{d}\to t\bar{b}$), $t$-channel ($ub\to dt$)
and $tW$ associated channel ($gb\to tW^{-}$); each process generates
distinct event distributions and can be measured separately. In the
LHT, the deviations of the single-top production rates of these three
processes from the SM predictions have to be identical at the tree
level, i.e. $\delta(s)=\delta(t)=\delta(tW)$. This is an important
test of the LHT. In contrast, the above relation does not hold for
LH models without T-parity.

\begin{table}

\caption{Mass limits (in unit of GeV) of $T_{+}$ and $T_{-}$ quarks for
various $\delta$ values. Here the brackets denote the range of $\delta$,
$M_{T_{+}}$ and $M_{T_{-}}$, respectively. \label{tab:mass-limie}}

\begin{tabular}{c|ccccc}
\hline 
$\delta\left(\%\right)$&
$\left[5,8\right]$&
~~&
$\left[2,5\right]$&
~~&
$<2$\tabularnewline
\hline 
$M_{T_{+}}$&
$\left[800,1000\right]$&
&
$\left[870,1600\right]$&
&
$>1100$\tabularnewline
$M_{T_{-}}$&
$\left[500,620\right]$&
&
$\left[580,950\right]$&
&
$>830$ \tabularnewline
\hline
\end{tabular}
\end{table}

\begin{figure}
\includegraphics[clip,scale=0.28]{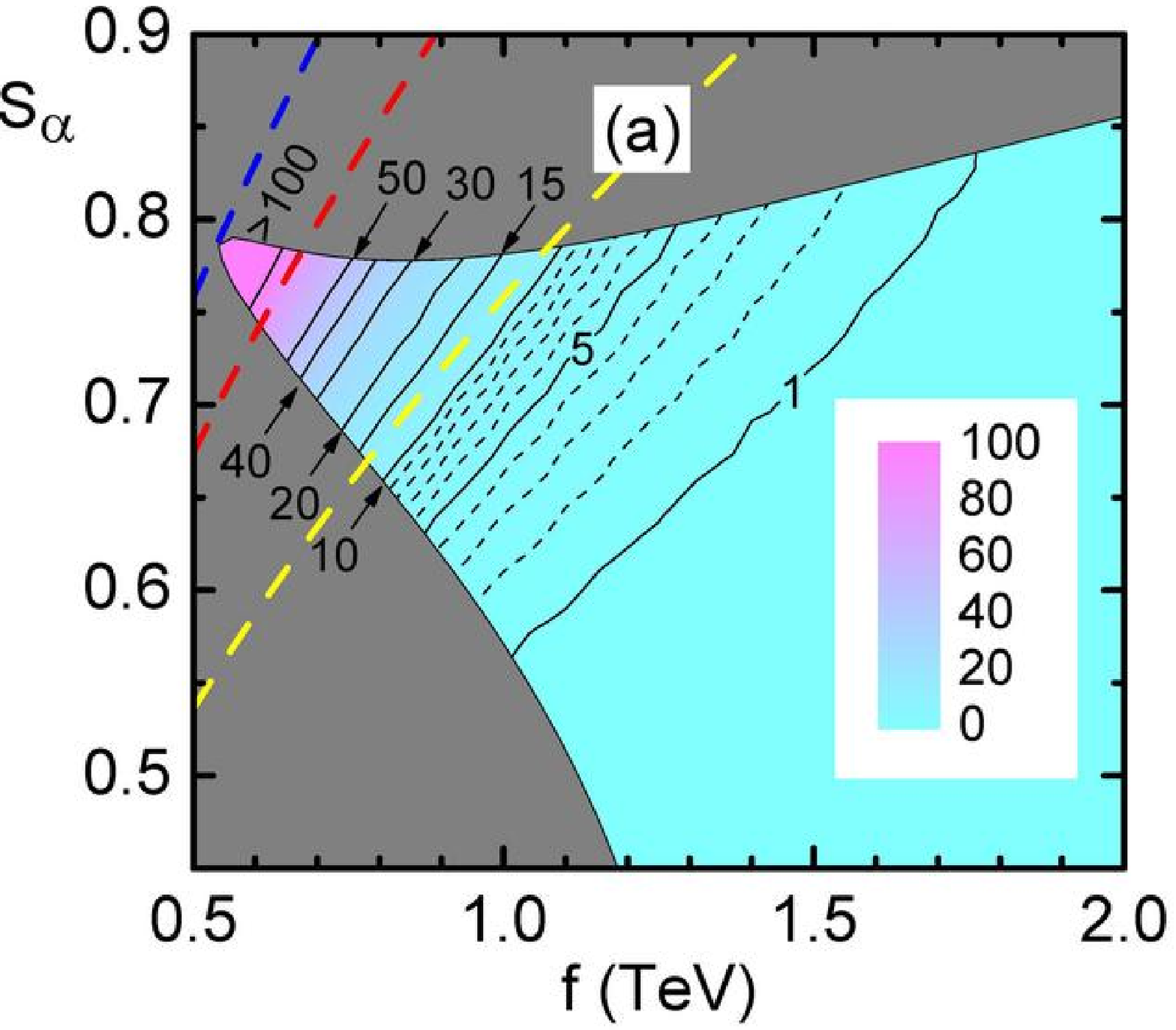}\includegraphics[clip,scale=0.28]{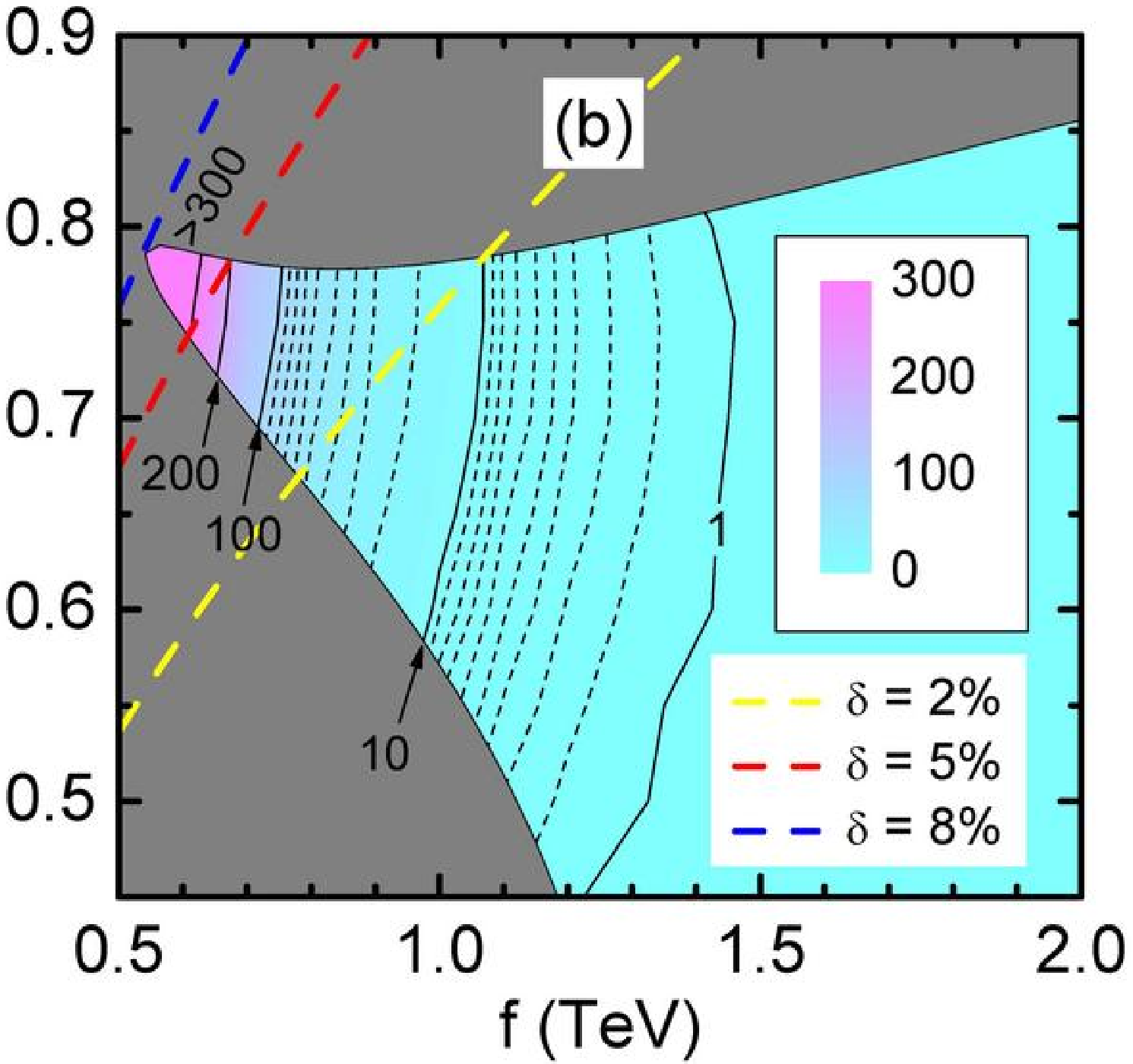}
\caption{Contours plots of the inclusive cross section (in unit of fb) in
the plane of $s_{\alpha}$ and $f$: (a ) for the single $T_{+}$
process given in Eq.~(\ref{eq:single-teven}) and (b) for the $\TT$
pair process given in Eq.~(\ref{eq:TTpari}). The $W$-boson decay
branching ratio is \emph{not} included here. (See detailed explanation
in the text.) \label{fig:xsec}}
\end{figure}

As noted above, the value of $s_{\alpha}$ cannot be zero in order
for the LH mechanism to take place. Therefore, a heavy $T_{+}$ can
be produced singly in hadron collisions via weak charged current ($W^{+}\bar{T}_{+}b$)
interaction, similar to the SM single-$t$ production, and is referred
as single-$T_{+}$ production in this work. In Fig.~\ref{fig:xsec}
we present the inclusive cross sections of single $T_{+}$ production
at the LHC in the plane of $s_{\alpha}$ and $f$. (Its production
rate is too small to be observed at the Tevatron for $f$ greater
than $500\,{\rm GeV}$.) In the same figure, we also show the constraints
from the single-$t$ production rate measurement on the parameters
$s_{\alpha}$ and $f$ for $\delta=2\%$ (yellow dashed-line), $5\%$
(red dashed-line) and $8\%$ (blue dashed-line), respectively. Again,
the gray region is excluded by EWPT. We note that the large single-$T_{+}$
cross sections ($\gtrsim50$\,fb) occur in the regime of $f<750\,{\rm GeV}$
and $s_{\alpha}\sim0.75$, where the single-top production rates are
reduced as compared to the SM rates by about $3\%\sim8\%$. Should
no deviation be found in the single-$t$ production, e.g. $\delta\leq2\%$
(below the yellow dashed curve), it will be very difficult to directly
observe the single-$T_{+}$ signal at the LHC due to its small cross
section ($\lesssim13$\,fb). Therefore, the correlation of the single
top process to the single-$T_{+}$ production can be used to test
the LHT.

While $T_{+}$ is mainly produced via single-$T_{+}$ process, the
T-odd heavy quark $T_{-}$ is predominantly produced in pairs via
strong QCD interaction because of the requirement of T-parity symmetry.
Since the coupling of gluon to $\bar{T}_{-}T_{-}$ pair is fixed by
the QCD gauge interaction, the $\TT$ pair production rate is determined
by the mass of $T_{-}$, and its dependence on the parameters $s_{\alpha}$
and $f$ is shown in Fig.~\ref{fig:xsec}. Again, we see that a precise
measurement of the single-$t$ production cross section can provide
a stringent test on the LHT. A reduction in the single-top production
rate by more than $5\%$ would imply the $\bar{T}_{-}T_{-}$ pair
production rate at the LHC to be larger than about $200\,{\rm fb}$
and $f$ to be less than about $700\,{\rm GeV}$. We also note that
the $\bar{T}_{-}T_{-}$ cross section is more sensitive to $f$ than
$s_{\alpha}$, as compared to the single-$T_{+}$ cross section.

The $T_{-}$ quark preferentially decays into $t$ plus $A_{H}$,
so the collider signature of the $\TT$ pair event is \begin{equation}
q\bar{q}(gg)\to T_{-}\bar{T}_{-}\to A_{H}A_{H}t\bar{t}\,,\label{eq:ttAA}\end{equation}
 where the two $A_{H}$'s produce the missing transverse energy signature.
For the $t\bar{t}$ plus missing transverse energy signature, the
intrinsic SM background is generated from the process $q\bar{q}(gg)\to t\bar{t}Z$,
where $Z$ decays into a pair of neutrinos, whose cross section is
about $60\,{\rm fb}$ at the LHC. To observe the $\TT$ signal, one
has to suppress this large background. Usually, this is done by making
kinematic selections to enhance the signal-to-background ratio, such
as the study done in Ref.~\cite{todd} which concluded that a large
background rate remained even after imposing a set of kinematic cuts.
Here, we propose a new method to largely suppress the SM background
rate by measuring the degree of polarization of the top quark (or
top anti-quark) in the final state.

\begin{figure}
\includegraphics[clip,scale=0.28]{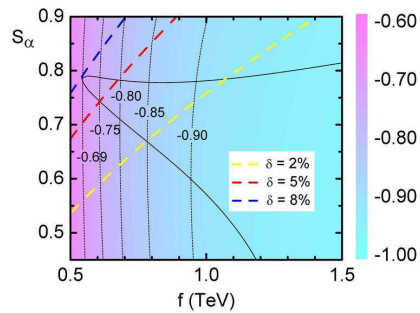}
\caption{Contour plot of $\mathcal{A}_{LR}$ in the $T_{-}\bar{T}_{-}$ pair
production in the plane of $s_{\alpha}$ and $f$. \label{fig:alr} }
\end{figure}

An interesting feature of the $T_{-}$ decay is that the top quark
from $T_{-}$ decay is predominately right-handedly polarized because
the left-handed component of the $A_{H}T_{-}t$ coupling is suppressed
by a factor of $\frac{v}{f}s_{\alpha}$, for \begin{equation}
g_{A_{H}T_{-}t}=\frac{2}{5}g^{\prime}s_{\alpha}\gamma_{\mu}\left(s_{\alpha}\frac{v}{f}P_{L}+P_{R}\right).\label{eq:coupling2}\end{equation}
 Parity is clearly broken in this coupling. In order to quantify the
parity violation effects, we define an asymmetry quantity $\mathcal{A}_{LR}$
as \begin{equation}
\mathcal{A}_{LR}\equiv\frac{\sigma(t_{L})-\sigma(t_{R})}{\sigma(t_{L})+\sigma(t_{R})}\,,\label{eq:alr}\end{equation}
 where $t_{L}$ and $(t_{R})$ denote the left-handedly and right-handedly
polarized top quark, respectively. The $Z$ boson preferentially couples
to a left-handedly polarized top quark such that $\mathcal{A}_{LR}^{SM}$
of the $t\bar{t}Z$ process is always larger than zero. (At the LHC,
$\mathcal{A}_{LR}^{SM}=0.106$.) On the contrary, $\mathcal{A}_{LR}^{LHT}$
of the signal process~(\ref{eq:ttAA}) is always smaller than zero.
Therefore, the SM intrinsic background rate can be largely suppressed
by demanding a negative value of $\mathcal{A}_{LR}$. In Fig.~\ref{fig:alr}
we present the contour of $\mathcal{A}_{LR}^{LHT}$ in the plane of
$s_{\alpha}$ and $f$. We note that $\mathcal{A}_{LR}$ mainly depends
on $f$ and is not sensitive to $s_{\alpha}$. This result leads to
an important observation that the new particle mass scale parameter
$f$ can be determined by measuring the asymmetry $\mathcal{A}_{LR}$
in the production rates of left-handed and right-handed top quarks
in the events with the $t\bar{t}$ plus missing transverse energy
signature. For example, if $\mathcal{A}_{LR}$ takes the value around
$-0.69$, then $f$ is about $550\,{\rm GeV}$. For a larger value
of $f$, the asymmetry $\mathcal{A}_{LR}$ approaches to $-1$. After
measuring $f$ via $\mathcal{A}_{LR}$, one can uniquely determine
$s_{\alpha}$ from the measurement of $M_{T_{-}}$ or $M_{T_{+}}$.
Should all these three observables ($\mathcal{A}_{LR},\, M_{T_{-}},\, M_{T_{+}}$)
be measured, together with the single-top precision measurements,
one can test the consistency of the LHT with data.

In the rest of the paper, we discuss how to directly measure $M_{T_{+}}$
and $M_{T_{-}}$ to test the LHT. $T_{+}$ quark has four tree level
decay channels which produce separately the $W^{+}b$, $Ht$, $Zt$
and $A_{H}T_{-}$ final states. Their decay branching ratios generally
depend on the model parameters $f$ and $s_{\alpha}$. In this study
we focus on the $T_{+}\to W^{+}b$ decay mode. The single-$T_{+}$
event could be detected via \begin{equation}
qb\to q^{\prime}T_{+}\to q^{\prime}bW^{+}\,,\label{eq:single-teven}\end{equation}
 where the $W$ boson decays leptonically. For the single-$T_{+}$
production, the SM backgrounds mainly come from the single-top processes
and the top quark pair production. The discovery potential of the
LHC for the single-$T_{+}$ production has been studied in the literature~\cite{teven}.
If the single-$t$ rate is found to be much smaller than the SM prediction,
then the LHT would predict a sizable single-$T_{+}$ signal which
is characterized by a much larger transverse mass (or scalar sum of
transverse energy) as compare to that predicted for the SM single-$t$
signal. The mass reconstruction of the $T_{+}$ quark is straightforward.
It is similar to the mass reconstruction of the single-top event.
One can first determine the longitudinal momentum of the neutrino
from the $W$-boson mass constraint and then reconstruct the ${T_{+}}$
invariant mass $M_{T_{+}}=\sqrt{\left(p_{b}+p_{\nu}+p_{\ell^{+}}\right)^{2}}$~\cite{neutrino}.

\begin{figure}
\includegraphics[clip,scale=0.6]{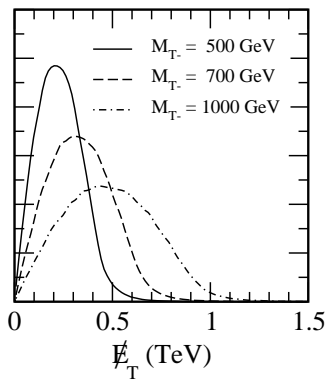}
\caption{Normalized distributions of $\met$ with different choices of $M_{T_{-}}$
for $s_{\alpha}=1/\sqrt{2}$. \label{fig:met-mtc} }
\end{figure}

One of the experimental signatures of the $\TT$ pair production at
the LHC is \begin{equation}
q\bar{q}(gg)\to T_{-}\bar{T}_{-}\to2A_{H}+t(\to bW^{+})+\bar{t}(\to\bar{b}W^{-})\,,\label{eq:TTpari}\end{equation}
 where $W^{+}$ decays leptonically and $W^{-}$ decays hadronically.
We have performed a Monte Carlo study and found a strong correlation
between the location of the (broad) peak of the missing transverse
energy (caused by the two LTP $A_{H}$ bosons) distribution in the
$\TT$ event and the mass of $T_{-}$. As shown in Fig.~\ref{fig:met-mtc},
the distribution of $\met$ peaks about half of $M_{T_{-}}$ for a
wide range of $f$ values. This special feature originates from the
spin correlation in the $\TT$ production.

In summary, we have shown that the measurement of the single-top production
is important for testing the LHT. Depending on the amount of its deviation
from the SM prediction, the masses of the heavy T-even ($T_{+}$)
and T-odd ($T_{-}$) partners of the top quark would be highly constrained.
Furthermore, the single-$T_{+}$ and the $\TT$ pair production rates
strongly depend on the result of the single-$t$ production cross
section measurement. We also proposed a new method to suppress the
SM background for detecting the $\TT$ events by noting that the signal
process tends to produce right-handed top quark from $T_{-}$ decay
while its SM background process ($t\bar{t}Z$) tends to produce left-handed
top quark. The asymmetry in the production rates of right-handed versus
left-handed top quarks in the events with $t\bar{t}$ plus $\met$
signature can be utilized not only to largely suppress the SM intrinsic
background, but also to provide a measurement of the $f$ parameter
itself. We also point out that because of the spin correlations in
the $\TT$ production and decay processes, the $\met$ distribution
peaks around half of the mass of $T_{-}$ which in turns provides
a new method for measuring $M_{T_{-}}$. From the measured values
of $f$ and $M_{T_{-}}$, one can determine the remaining parameter
$s_{\alpha}$.

\textbf{Acknowledgments} We thank A.\ Belyaev, C.-R.\ Chen and K.\
Tobe for discussions. Q.H.C. and C.P.Y. are supported in part by the
U.S. DOE and NSF under grant No. DE-FG03-94ER40837 and award No. PHY-0555545,
respectively. C.S.L. is supported by the China NSF under grant No.
10421503 and No.10575001.

\end{document}